\documentclass[
showpacs,
preprint,
]{revtex4}

\usepackage{graphicx}
\usepackage{bm}
\usepackage{graphicx}

\usepackage{color}

\begin{document}

\title{A reduction in the UHE neutrino flux due to neutrino spin precession}

\author{J. Barranco$^1$}
\email{jbarranc@fisica.ugto.mx}

\author{O. G. Miranda$^2$}
\email{omr@fis.cinvestav.mx}  

\author{C. A. Moura$^3$}
\email{celio.moura@ufabc.edu.br}

\author{A. Parada$^2$}
\email{aparada@fis.cinvestav.mx}

\affiliation{$^1$Divisi\'on de Ciencias e Ingenier\'ias,  Universidad de
Guanajuato, Campus Le\'on, C.P. 37150, Le\'on, Guanajuato, Mexico}

\affiliation{$^2$Departamento de F\'{\i}sica, Centro de Investigaci{\'o}n y de
  Estudios Avanzados del IPN, Apdo. Postal 14-740 07000
  M\'exico, D.F., Mexico}

\affiliation{$^3$Centro de Ci\^encias Naturais e Humanas, Universidade
  Federal do ABC, Rua Santa Ad\'elia, 166, 09210-170 Santo Andr\'e,
  SP, Brazil}

\date{\today}
\begin{abstract}
Motivated by the stringent flux limits for UHE neutrinos coming from
gamma ray bursts or active galactic nuclei, we explore the possibility
that the active neutrinos generated in such astrophysical objects
could oscillate to sterile right handed states due to a neutrino
magnetic moment $\mu_\nu$. We find that a value as small as
$\mu_\nu\approx 10^{-15}\mu_{\rm B}$ could produce such a transition
thanks to the intense magnetic fields that are expected in these
objects. \\
Keywords: Cosmic rays, Neutrinos, Neutrino electromagnetic properties.
\end{abstract}
\maketitle

In recent years the observation of very distant astrophysical sources,
such as Active Galactic Nuclei (AGN) and Gamma Ray Bursts (GRB), have
improved notoriously. Now we have a better knowledge of these objects,
despite the fact that there is still a lot of puzzles to be
unraveled. It is a general belief that GRBs and AGNs provide a
mechanism for the acceleration of the most energetic cosmic rays that
have been detected so far. One of the reasons for this belief are the
strong magnetic fields inside these objects, that may accelerate
protons and heavier nuclei up to the highest energy range of the
spectrum of the cosmic
radiation~\cite{Hillas:1985is,Torres:2004hk}. Currently, cosmic rays
with energies as high as $10^{20}$~eV have been detected at different
experiments on Earth. However, there is a limit for protons with
energies above $10^{20}$~eV to travel distances larger than
100~Mpc~\cite{Greisen:1966jv} and, therefore, there is no physical
chance to obtain direct information about the most distant sources
from ultra-high energy protons, as has been confirmed by
HiRes~\cite{hires} and the Pierre Auger
Observatory~\cite{Abraham:2008ru}.

In principle, this limitation do not apply to neutrinos and it would
be expected that, at energies around $10^{18}$~eV it would be possible
to observe neutrinos coming from extragalactic sources and obtain, at
least in principle, direct information from their original
source. This has been one of the main motivations of the IceCube
experiment~\cite{Resconi:2008fe}. Recent reports from several
experiments, however, show negative results in the search for
extragalactic neutrinos, giving upper limits for a diffuse or for
point source neutrino
fluxes~\cite{Abbasi:2012zw,Guardincerri:2011zz,Biagi:2011kg,Abbasi:2008hr}.
Although it could be possible that in future, with more statistics, a
positive detection and a determination of the UHE neutrino flux could
be established, we think that it is a good time to search for possible
alternative explanations to the absence or reduction in the flux of
neutrinos.  There have been efforts to understand effects on the
neutrino flux and other cosmological observables, such as the CMB
power spectrum, by considering an interaction of the neutrino with a
Dark Matter candidate. However, most of these attempts lead to small
effects~\cite{Weiler:1983xx,Weiler:1992fm,Roulet:1992pz,Mangano:2006mp,Boehm:2006mi,Boehm:2004}.
The main problem to explain a suppression in the neutrino flux due to
this kind of interaction is that the neutrino-Dark Matter expected
cross section is too small to play an important role, except for the
case of an ultra-light scalar field ($m_\phi \sim
10^{-23}-10^{-33}$~eV) where the small cross section is compensated by
the large amount of DM particles~\cite{Barranco:2010xt}.

Here we focus on a different approach that may be  simpler and
physically appealing: the case of a spin flip of the neutrino due to
a non zero neutrino magnetic moment. In the Standard Model (SM), the
neutrino magnetic moment is expected to be extremely
small~\cite{Vogel:1989iv,Lee:1977tib,Marciano:1977wx}:
\begin{equation}
\mu_\nu=\frac{3G_F m_e m_\nu}{4\sqrt{2}\pi^2} 
  = 3.2\times10^{-19}\left( \frac{m_\nu}{\rm [eV]}\mu_B  \right) .
 \end{equation}
However, motivated by the solar neutrino problem, it was noticed that
a relatively large neutrino magnetic moment could play a role in
neutrino conversion inside the Sun. The most successful mechanism in
this direction was the well known Resonant Spin Flavor Precession
(RSFP)~\cite{Akhmedov:1988uk} where an oscillation
$\nu_e\to\bar{\nu}_{\mu ,\tau}$ may occur. 
Despite RSFP was not able to be a
solution of the solar neutrino problem~\cite{Barranco:2002te} it
motivated several theoretical efforts to construct models beyond the
SM that could explain a large value of the neutrino magnetic
moment~\cite{Babu:1990hu,Babu:1989wn,Barbieri:1988fh}. 
This same mechanism had
also been applied in the past to the case of UHE cosmic
rays~\cite{Enqvist:1998un,Sahu:1998bh}, mainly motivated by the
possibility to detect tau neutrinos appearing from oscillation during
the neutrino propagation in cosmological
distances~\cite{Athar:2005wg}. Tau neutrinos could be identified
by very unique signatures such as double bang
events~\cite{Learned:1994wg,Moura:2007zz} and Earth
skimming~\cite{Feng:2001ue}. 
At present, thanks to a remarkable experimental effort, there are
limits to the neutrino magnetic moment as strong as $\mu_\nu \leq
10^{-11}\mu_B$ coming from laboratory
measurement~\cite{Wong:2006nx,Daraktchieva:2005kn} or from a combined
analysis~\cite{Grimus:2002vb}, and $\mu_\nu \leq 10^{-12}\mu_B$ from
astrophysical observations~\cite{Raffelt:1990pj} or from solar
data~\cite{Miranda:2003yh}.

Given the non observation until now of UHE neutrinos, our main goal in
the study of spin flip neutrino conversion due to a neutrino magnetic
moment will be the transition from an active electron neutrino
(presumably produced in an extragalactic object such as an AGN or a
GRB) into a right handed sterile neutrino. Such a conversion may take
place in different scenarios. We can consider in the first place the
case of a conversion due to a diagonal magnetic moment that converts
the active electron neutrino into a right handed sterile electron
neutrino. This case had been considered as a possible explanation of
the solar neutrino problem long time ago~\cite{Cisneros:1970nq} and
the conversion probability in this case is given by
\begin{equation}
P(\nu_{e_L}\to\nu_{e_R};r) = \sin^2\left( \int_0^r \mu_\nu B_\perp(r')dr'\right) .
\label{eq:cisneros}
\end{equation}

As has been noticed before~\cite{Akhmedov:1997yv,Athar:1995cx}
  there is a possibility in this picture that a neutrino flux can be
  fully converted into sterile neutrinos if the condition~\footnote{for a
  constant magnetic field}
\begin{equation}
\mu_\nu B_\perp r\approx \frac{\pi}{2} \label{eq:condition}
\end{equation}
is satisfied. 

Considering the vast range of both magnetic field intensities and
sizes of the astrophysical objects it would be not a surprise that,
for a reasonable value of the neutrino magnetic moment, there will be
astrophysical objects that could induce a spin conversion, while for
an unsuitable combination of this values the effect will not be valid
(otherwise this will be a fine tuning).

Therefore, if future experimental results continue reporting no
observation of neutrinos for certain objects, or for certain neutrino
flavors, this could be a clue for a Dirac magnetic moment. On the
other hand the future experimental results should give, at the same
time, positive neutrino signals for astrophysical objects that do not
fulfill the requirements of Eq.~(\ref{eq:condition}).

\begin{figure}
\includegraphics[width=0.7\columnwidth]{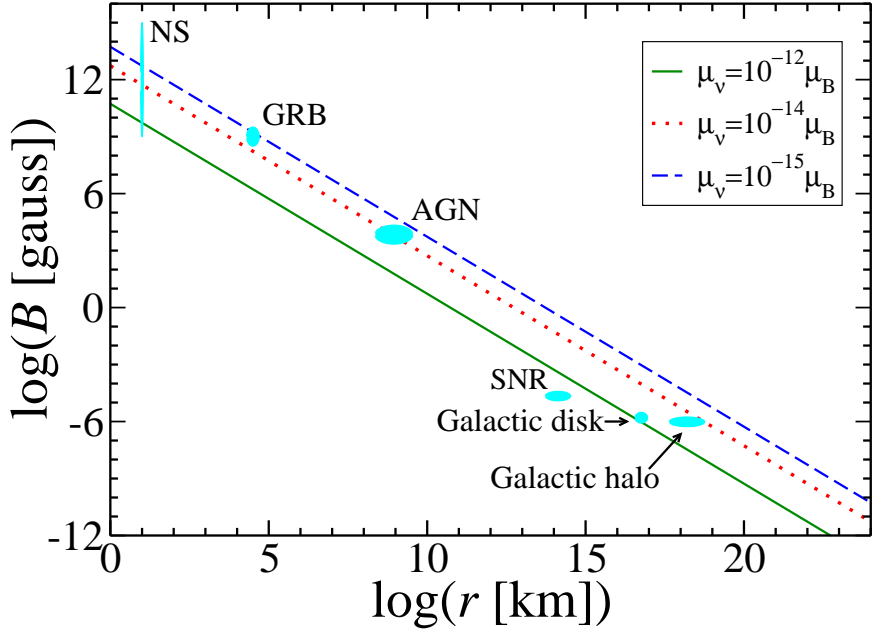}
\caption{ 
Relation of magnetic field $B$ and size $r$ of astrophysical
  sources for an efficient neutrino spin transition
  $\nu_{e_L}\to\nu_{e_R}$. The  curves show different values
  of the neutrino magnetic moment. The acronyms refer NS for 
  Neutron Stars, GRB for Gamma Ray Bursts, AGN for Active 
  Galactic Nuclei, and SNR for Supernova Remnants.
}\label{fig:newhillas}
\end{figure}

We show in Fig.~\ref{fig:newhillas} the regions in the $B-r$ plane
that satisfy the above condition for a diagonal neutrino magnetic
moment of $10^{-12}\mu_B$,~$10^{-14}\mu_B$, and $10^{-15}\mu_B$.
Inspired by the Hillas Plot~\cite{Hillas:1985is}, we also show in the
same figure the astrophysical objects that lie in such regions. For a
given neutrino magnetic moment, the astrophysical objects lying in the
corresponding curve may induce a neutrino transition into a sterile
state.  In this picture, a relatively small neutrino magnetic moment,
e.g., of the order of $\mu_\nu=10^{-15}\mu_B$, could produce an
efficient conversion into sterile states in the case of GRB, an
interesting feature considering the recent limit for the neutrino flux
coming from such objects~\cite{Abbasi:2012zw}. Note that a higher
neutrino magnetic moment around~$10^{-14}\mu_B$, could induce the same
effect for an AGN; in this case, there could be a very efficient
mechanism for the suppression of neutrinos coming from the AGN, 
  since the condition of Eq. (\ref{eq:condition}) would be satisfied,
  while the flux for a GRB would only be suppressed by a factor one
  half due to the high value of $\mu_\nu$; as mentioned above, in this
  case a future positive signal of GRB neutrinos combined with a
  negative result for AGN could be a hint for a nonzero neutrino
  magnetic moment of the order of~$10^{-14}\mu_B$.  Note also that, at
  least in first approximation, the weak magnetic field in the
  galactic halo and intergalactic medium may also produce a spin
  conversion given the long distance traveled by the neutrino flux.

We believe that, given the fact that there has been no observation of
neutrinos coming from AGNs or GRBs, it would be important to
consider this mechanism in more detail. Besides the detailed
comparison with the experimental results, it would also be important
to consider matter effects~\cite{Okun:1986na,Okun:1986hi}, that might
diminish the mechanism. For constant density matter the conversion
probability in this case will be given
by~\cite{Akhmedov:1997yv,Okun:1986na,Okun:1986hi,Barbieri:1987xm}
\begin{equation}
P = \frac{(2\mu_\nu B_\perp)^2}{ V_e^2 + (2\mu_\nu B_\perp)^2 } \sin^2\left(
\frac12 \sqrt{V_e^2 + (2\mu_\nu B_\perp)^2}\,r\right) .
\label{eq:vvo}
\end{equation}
With $V_e= \sqrt{2}G_F(N_e-N_n/2)$, $G_F$ the Fermi constant, and
$N_{e,n}$ the electron and neutron densities.  It is possible to see
from this formula that a high value of the potential $V_e$ would
suppress the spin conversion. This is not the case for an AGN or a
GRB. We show in Table \ref{table:Ve} the approximate expected values
of the potential, considering only the $N_e$ contribution, and compare
them with the product of the neutrino magnetic moment and the expected
magnetic field strength at the source. One can see that the potential
is always negligible.

\begin{table}
 \begin{tabular}{lrrr}
\hline \hline 
Source & $V_e$ (eV) &  &  $\mu B$ (eV) \\ 
\hline\hline 
GRBs~\cite{Chevalier:1999jy} & $2\times 10^{-34}$ & & $10^{-13}$ \\ 
AGNs~\cite{AlvarezMuniz:2004uz} & $10^{-27}$&  & $6 \times 10^{-20}$\\ 
SNRs~\cite{Xu:2009vc}        & $ 10^{-37}$ & & $  10^{-28}$ \\ 
Galactic Disk~\cite{deAvillez:2012ue} & $5\times 10^{-39}$&  & $5 \times 10^{-29}$
\\ \hline\hline
\end{tabular}
\caption{Comparison of the expected electron density versus the
  product of a $10^{-14}\mu_B$ neutrino magnetic moment times the
  magnetic field of the astrophysical source. }
\label{table:Ve}
\end{table}

Another important mechanism to consider would be an {\it spin flavor}
precession into a different sterile neutrino flavor. In this last case
we consider the evolution equation

\begin{equation}
i\left(
\begin{array}{l}
\dot{\nu}_{e_L}\\
\dot{\nu}_{x_R}\\
\end{array}
\right) =
\left(
\begin{array}{cc}
V_e - \delta &  \mu_\nu B_{+}  \\
\mu_\nu B_{-} &  \delta  \\
\end{array}
\right)
\left(
\begin{array}{c}
\nu_{e_L}\\
\nu_{x_R} \\
\end{array}
\right)\,,
\label{master}
\end{equation}
where $\mu_\nu$ denotes now a neutrino transition magnetic
moment~\cite{Schechter:1981hw}, $B_\pm=B_x\pm iB_y$ and $\delta =
(\Delta m^2/4E_\nu) \cos2\theta$ with $\Delta m^2$ the neutrino mass
difference parameter, $\theta$ the corresponding neutrino mixing angle
and $E_\nu$ the neutrino energy.  Finally, $x$ may denote a $\mu$ or
$\tau$ neutrino or even a new sterile state, in which case we are not
constrained to the squared mass differences of the active neutrino
states and, therefore, we could have more room to consider a sterile
neutrino even in the range of keV.  However, it is important to note
that in this case the conversion probability will depend also on the
mass square difference~\cite{Akhmedov:1997yv,Akhmedov:1988uk}:
\begin{eqnarray}
P_{\nu_{eL}\to\nu_{xR}} &=& \frac{(2\mu_\nu B_\perp)^2}{
(2\delta - V_e)^2 + (2\mu_\nu B_\perp)^2
}  \nonumber \\
&\times&  \sin^2\left( \frac12 \sqrt{(2\delta -V_e)^2 + (2\mu_\nu B_\perp)^2}\,r\right)\,.
\label{eq:rsfp}
\end{eqnarray}
From this expression, and comparing for the case of GRBs
($E_\nu\approx 10^{15}$~eV) or AGNs ($E_\nu\approx 10^{18} $~eV), it
is possible to see that even in the case of the standard neutrino mass
differences ($\Delta m^2_{13}= 7.6\times 10^{-5}$~eV$^2$ and $\Delta
m^2_{23}= 2.5\times 10^{-3}$~eV$^2$~\cite{Tortola:2012}) the value of
$\delta$ gets closer to the product $\mu_\nu B_\perp$ and,
consequently, it is in the limit to suppress the conversion mechanism,
while a conversion into a heavier sterile neutrino, such as a keV
neutrino, will be certainly suppressed. Finally, considering a random
magnetic field, instead of the regular case that we have discussed, is
of no help in this case since it has been shown that in this case the conversion
probability into a sterile state is at most of one
half~\cite{Domokos:1997cq}.

In the present day there are several experiments, like IceCube and the
Auger Observatory, expecting to detect extragalactic neutrinos. Until
now, neutrinos with energy above $10^{15}$~eV, coming from
extragalactic sources, have not been detected.  Based on the non
observation of extragalactic neutrinos,
AGN~\cite{AlvarezMuniz:2004uz,Mannheim:1998wp,Halzen:1997hw,
  Mannheim:1995mm,Stecker:1995th,Stecker:1991vm} and
GRB~\cite{Ahlers:2011jj,Murase:2007yt,Rachen:1998ir,Waxman:1997ti}
models that predict high observable neutrino fluxes could be excluded,
but with the mechanism of neutrino flavor conversion that we have
discussed, this apparent contradiction may not exist. The neutrinos
could be generated in the sources but converted into sterile neutrinos
due to the strong magnetic fields that prevails in those environments.

In this work we have stressed the possibility of an efficient
transition of the neutrinos into a right handed sterile neutrinos due
to a non zero magnetic moment and due to the presence of strong
magnetic fields both in GRBs as well as in AGNs. We consider this is
an interesting mechanism that could be studied in more detail as more
experimental results appear. If the current tendency of getting strong
limits on the UHE neutrino flux continues, this could be a hint for a
non zero neutrino magnetic moment effect, while a positive observation
could put a stronger limit on $\mu_\nu$. Moreover, in this picture it
would be natural that different objects could produce different
reduction rates, providing a way to test the mechanism if future
experimental results could detect neutrinos from different sources.

Note: After the first version of this work, other articles have
  discussed different mechanisms that could also lead to a neutrino flux
  suppression~\cite{Baerwald:2012at,Esmaili:2012fu}. Besides, they also
  discussed the recent claim of a possible detection of electron
  neutrinos by IceCube (while no muon neutrinos have been yet
  detected)~\cite{Ishihara}. We would like to note that, if these were
  the case, the mechanism discussed here could also work, for instance, with 
  a vanishing neutrino magnetic moment for electron neutrinos and, for
  $\nu_\mu$ case, a magnetic moment value of the order discussed
  above. 

\begin{acknowledgments}
  This work has been supported by CONACyT grant 132197 and SNI-Mexico.
  A. P. has been supported by CONACyT grant 166639.  C.A.M. thanks the
  Physics Department of CINVESTAV for the hospitality during the visit
  when part of this work was done.
\end{acknowledgments}

\end{document}